# THE DEMAND SIDE OF OPEN GOVERNMENT DATA: A CASE STUDY OF KINGDOM OF BAHRAIN

Abdulkarim Katbi[1], Jaflah AlAmmari[2], Ali AlSoufi[3]

[1]King Hamad University Hospital, Kingdom of Bahrain
abdokatibi@yahoo.com
[2]University of Bahrain, Kingdom of Bahrain
jalammari@uob.edu.bh
[3]British University of Bahrain, Kingdom of Bahrain
A.AlSoufi@BUB.BH

## ABSTRACT

*Governments around the world have realized the importance of Open Government Data - OGD as new paradigm shift in government that focuses on making governments more service oriented, transparent and competent. However, as with many countries, the situation of OGD initiative in Kingdom of Bahrain is not promising as reflected by number of assessments that measure the implementation and progress of OGD worldwide. The current research aims at investing the local situation regarding consuming and reusing OGD in Kingdom of Bahrain. Specifically, this research assesses the level of citizen awareness towards OGD, determines citizens' requirements of OGD and identifies the key challenges and obstacles in using/reusing OGD. A questionnaire was developed to investigate the demand side of OGD. The findings show that serious and responsible efforts from the publishers of OGD, namely: Government Organizations are believed to be a necessity in order to progress the implementation process of OGD initiative in Kingdom of Bahrain.*

## KEYWORDS

*Open Government Data, electronic government, supply, demand, digital society, Kingdom of Bahrain*

## 1. INTRODUCTION

Open government data is relatively a recent topic that are inspired by many changes including the worldwide movement of governments towards embracing more open and transparent theme to aspire citizens to innovate and become more involved with their respective governments. Starting from 2009, several countries have launched an open data initiative and joined the Open Government Partnership to show their commitment towards opening governmental data. Examples of the early adopters of OGD include USA, England and Australia[1], [2].

The new advancements in information technologies have made the realization of Open Government data concept possible through different means. For civilians, their ability to access, manipulate and share the data was enabled in an unprecedented way by the digital technologies. In fact, the digital society became an integral part of  made more possible by newer technologies.

The value creation of Open government data is largely dependent on the actions of the governments towards the publication of OGD. Since governments are the main collectors of public data, selling or hiding the data from the public will not help in fostering the OGD initiative. In fact, the benefits of OGD cannot be reaped unless the data is made available to the public for their unrestricted use [3].

The complexity and the unique nature of OGD requires paying attention to multiple influencing constructs in the OGD domain. Many OGD lifecycle and ecosystem approaches were proposed to outline the broader picture of OGD involving the key steps and processes, key stakeholders and the various dependencies among them [4]. Notably, all OGD lifecycle and ecosystem approaches implicitly or explicitly refer to the demand side of OGD as an essential aspect in the OGD domain [5][6][7][8][9].

Assessment approaches are necessarily for successful embedding of OGD initiative in society and improving the uptake of released data by the various stakeholders [7]. This is because of the fact that Open government data initiatives usually require the involvement and commitment of various stakeholders, considerable investments as well as efforts to implement, monitor and improve the initiatives [5]. Thus, assessment approaches help in identifying how governments can effectively engage with society as a whole to stimulate use/reuse of OGD, determine the actions required to improve the quality and usability aspects of published data and possibly strengthening the OGD ecosystem to deliver social, economic and political impacts [10].

One of the main and current issues with OGD is that the assessments and evaluations are being directed mainly towards the supply side of OGD [7]. In particular, scholars and practitioners are paying more attention for evaluating aspects such as the production of the data and validating the qualitative quality dimensions of the portal, data and metadata produced [11][12][13][14]. While such aspects are necessary, studying the demand side of OGD is of utmost importance as well to ensure successful implementation of the initiative.

Within the context of Kingdom of Bahrain, both practitioners and scholars affirm the need for more efforts to improve the OGD initiative. Practitioners stated clearly that Kingdom of Bahrain is not at the top performers' list with respect to OGD implementation and use [15] [17]. Scholars raised concerns regarding the limited OGD research in kingdom of Bahrain and particularly in the demand side of the initiative [18][19] [20]. The economic vision 2030 [21] and nation-wide smart city initiative in Kingdom of Bahrain both necessitates studying the demand side and determining what actions are required to improve the OGD initiative.

Since OGD initiative in Kingdom of Bahrain still lacks many factors to make it successful, studying the demand side of it is of vital importance. The identification of users' requirements, needs and challenges with respect to OGD will enable more effective implementation and gaining more benefits of OGD initiative. A successful implementation would contribute directly to the economy of the country. It is expected to gain around 1% direct increase in GDP out of a mature OGD initiative, as reported by [22]. Thus, the economic vision of 2030 will be supported by the mature OGD initiative. Additionally, the government would be more able to prioritize the corrective and implementation actions of OGD based on the requirements of the users. This enables the movement towards creating an environment in which the government becomes more service oriented, transparent and competent (Gov 3.0)[23] . Smart city projects will also be benefited from evaluating the demand side of OGD initiative. Since OGD users comprise the business sector workers, expressing and sharing their demands and concerns will help the government officials to focus on their needs and try to fulfill them accordingly.

This paper attempts to evaluate the demand side of OGD in kingdom of Bahrain by following two main activities. First, a structured literature review is conducted to identify the important elements related to the demand side of OGD. Second, a survey-based quantitative research methodology is adopted to collect and analyze the data. Accordingly, the paper is organized as follows. Section 2 presents a literature review related to OGD. The research methodology is presented in section 3. Section 4 presents the results of the study. A detailed discussion of the results is presented in section 5. Finally, section 6 concludes the study with recommendations.

## 2. LITERATURE REVIEW

Open government data refers to structured, machine-readable and machine-actionable data that governments and publicly funded research organizations actively publish on the internet for public reuse and that can be accessed without restrictions and used without payment [24]. Open government data initiative can be defined as set of activities being undertaken by the government bodies at different levels to publish open data, promote the use /reuse of it, monitor and evaluates the progress and take corrective actions whenever necessary to insure successful implementation [25]. To this extend, Open government data initiatives are considered as an extension or a subset of e-government [26].

Several researchers and practitioners have demonstrated the benefits of OGD. Attard et al.[5] showed three main benefits that are believed to be essential for motivating governments to embrace OGD. The benefits are increased transparency, releasing social and economic value and participatory governance. Martin and Begany [27] studied the benefits of opening specific type of government data (health data) to the public. Benefits that could be gained include more efficient public health operations, improved healthcare delivery and health literacy, and reaching new audiences. Zuiderwijk et al.[28] did a comprehensive review of the benefits mentioned in literature and policy documents. Other studies and assessments that considered the benefits of OGD include [29][25][30].

Despite the benefits of OGD, adoption barriers, challenges and obstacles also exist and could prevent the OGD initiative from reaching its full potential. Zuiderwijk et al. [31] identified eight categories of barriers. The identified barriers are availability and access, findability, usability, understandability, quality, linking and combining data, comparability and compatibility, and metadata impediments. Janssen et al. [30] acknowledged a list of adoption barriers for OGD including institutional, task complexity, use and participation, legislation, information quality, and technical barriers. Zuiderwijk and Janssen [32] examined the open data lifecycle and identified barriers related to the use and publication of open data under each process of OGD lifecycle.

Barriers related to specific groups or private companies were also investigated. Brugger et al. [33] examined the barriers affecting the usage of OGD by media, political parties, associations and Non-governmental organizations (NGOs). Chorley [34] surveyed the records management team of England healthcare institutions to identify the challenges they face because of implementing the OGD processes. Magalhães and Roseira [35] explored the barriers that private companies encounter when using OGD. Other OGD-related barriers were investigated by [5] and [36].

OGD ecosystem approaches presented in literature confirm the importance of considering the users of OGD. For example, Donker and Loenen [7] adopted an open data ecosystem thinking to illustrate how existing assessment approaches focuses only on certain aspects and ignoring other important elements of the bigger OGD ecosystem -like demand side of OGD. Their ecosystem consists of five main constructs, namely Data (supply side), Use (demand side), access networks, policy and standards. Dawes et al.[9] provided a general OGD ecosystem model that focuses on elements such as policy and strategy, the general settings, OGD publication and use, benefit generation, and feedback and communication. Najafabadi and Luna-Reyes [37] proposed a detailed OGD ecosystem to enhance the understanding of OGD enablers and barriers and assist policymakers to act towards improving the OGD initiative. Their model consisted of four main layers, namely Data, applications, use and benefits. Other ecosystem approaches include [8] [38][39][40] .

There exist a limited attention and focus of scholars to study and evaluate the OGD initiative in Kingdom of Bahrain. Elbadwi [19] reviewed the OGD Portal based on metrics such as data formats, metadata, data set format, presentation and quantity, Participation and collaboration mechanisms. The results of his study doubted the usability of Bahraini's OGD portal and confirmed the lack of documented and communicated strategy that focuses on OGD initiative at a national context. Alanazi and Chatfield [41] measured the maturity of OGD across Middle East countries based on the eight principles of OGD identified by open government working group [42]. The results demonstrated some evidence that the surveyed datasets have complied with the basic principles of OGD openness. Alromaih et al.[43] examined OGD portals of GCC countries through proposing a list of technical features and evaluating the conformance of the portals to the list. The results confirmed that OGD portal of Bahrain is in the second least technically mature portal in GCC after Saudi Arabia.

A more geographical spanned assessment was carried out by Martín et al.[13]. In their assessment, they examined and compared between the quality of OGD portals provided by national governments worldwide. The results of the study conclude that Bahrain's OGD portal still lacks many essential aspects to qualify it as a good and robust OGD portal. Despite the presence of functional features, both semantic and content aspects of the OGD portal were found to be lacking and insufficient.

Despites the technical considerations of the OGD initiative, Saxena [20][44] explicitly asserted that the demand side of OGD initiative should not be neglected. During her assessment process, Saxena [20] stated that the OGD initiative in GCC countries, including Kingdom of Bahrain, are still not mature. The way of OGD initiative implementation in GCC countries exhibited a unidirectional method of suppling the data from the government to the public. Saxena [20] claimed that OGD implementation can be hindered if the involvement of users is not considered. Another research by Saxena [44] suggests that the implementation potential of OGD initiative is high in Kingdom of Bahrain, but the use potential among citizens is low. With this regard, Saxena [44] stated that despite having a good infrastructure, the government has to take actions to stimulate citizen participation through increasing OGD awareness among them as well as try to benefit from the best practices of OGD leaders worldwide.

## 3. RESEARCH METHODOLOGY

### 3.1 Research Question and objectives

In order to establish a well-defined approach for the current research paper, it is essential to identify the overall research question and the objectives of the study. Since the goal of this survey is to get more insights regarding the demand side of OGD initiative in kingdom of Bahrain, the following research question has been defined:

*What is the current situation regarding using open government data in Kingdom of Bahrain?*

In order to assist in finding answers to the research question, number of objectives were identified as follows:

- Assess the level of people's awareness and use towards OGD.
- Determine people's requirements of OGD.
- Identify the people's perceived benefits and barriers of OGD

### 3.2 Research methodology and data collection approach

With respect to the current study, a quantitative research approach was selected. Since the current study focuses on the users of OGD, the targeted population are the people living in kingdom of Bahrain (potential OGD users). As the total number of population is known and it is possible for the researcher to give the citizens "an equal chance of being selected in relation to their proportion within the total population" [45], probability sampling was adopted. Probability sampling allows the research objectives to be achieved by statistically estimating the characteristics of the population from a sample [46].

With regards to the sampling technique, it is necessary to adopt a sampling technique that would cover all the Four governorates in the Kingdom; the Capital, Muharraq, Northern and Southern. Therefore, a modified version of cluster sampling was adopted in which the population will be divided first into several clusters – the four governorates and then random sampling will be applied for each and every cluster to collect the data. Table 1 shows the total number of people living in each cluster (governorate) and the percentage to the overall population in Kingdom of Bahrain. 95% confidence level and a 5% margin of error (confidence interval) were adopted in order to calculate the sample size. Table 2 shows the distribution of the calculated sample over for governorates based on the percentage to the total population identified earlier.

Table 1 Distribution of citizens per Governorate

| Cluster Name | Capital | Muharraq | Northern | Southern | Total |
|---|---|---|---|---|---|
| Number of samples required | 148 | 67 | 92 | 77 | 384 |

Table 2 Distribution of the calculated sample over governorates in Kingdom of Bahrain

| Cluster Name | Capital | Muharraq | Northern | Southern | Total |
|---|---|---|---|---|---|
| Population | 577,432 | 263,201 | 359,425 | 301,058 | 1,501,116 |
| Percentage to the total population | 38.47 | 17.53 | 23.94 | 20.06 | 100 % |

## 4. DATA ANALYSIS AND RESULTS

In order to get an overview of demographic dispersion of individuals living in Kingdom of Bahrain, general factors such as age, gender, educational Level and working sector have been displayed in the following. Results in Table 3 show that 60% women and 40% men participated in the survey and displayed a relatively close to equal dispersion of both genders in the evaluation of findings. The majority of participants were in the age of 21 to 40 years. With respect to educational level, almost half of respondents holds a bachelor's degree, one quarter of them holds a diploma and 30% of them holds a high school or below degree. Individuals who hold a post-graduate degree were the least and

| | | Percent |
|---|---|---|
| **Gender** | Male | 39.8 |
| | Female | **60.2** |
| | Total | 100 |
| **Age (years)** | 20 or under | 14.1 |
| | 21-30 | **51.1** |
| | 31-40 | 25.4 |
| | 41-50 | 6 |
| | 51+ | 3.5 |
| | Total | 100 |
| **Educational Level** | High School or below | 28.5 |
| | Diploma | 14.8 |
| | Bachelor's degree | **49.3** |
| | Post-graduate degree | 7.4 |
| | Total | 100 |

accounted for only 7.5% out of the total population. With respect to the work sector,

shows that almost half of respondents work in private sector followed by 24% percent of respondent working in the public sector.

Table 3 Respondents demographic data

| | | Percent |
|---|---|---|
| **Gender** | Male | 39.8 |
| | Female | **60.2** |
| | Total | 100 |
| **Age (years)** | 20 or under | 14.1 |
| | 21-30 | **51.1** |
| | 31-40 | 25.4 |
| | 41-50 | 6 |
| | 51+ | 3.5 |
| | Total | 100 |
| **Educational Level** | High School or below | 28.5 |
| | Diploma | 14.8 |
| | Bachelor's degree | **49.3** |
| | Post-graduate degree | 7.4 |
| | Total | 100 |

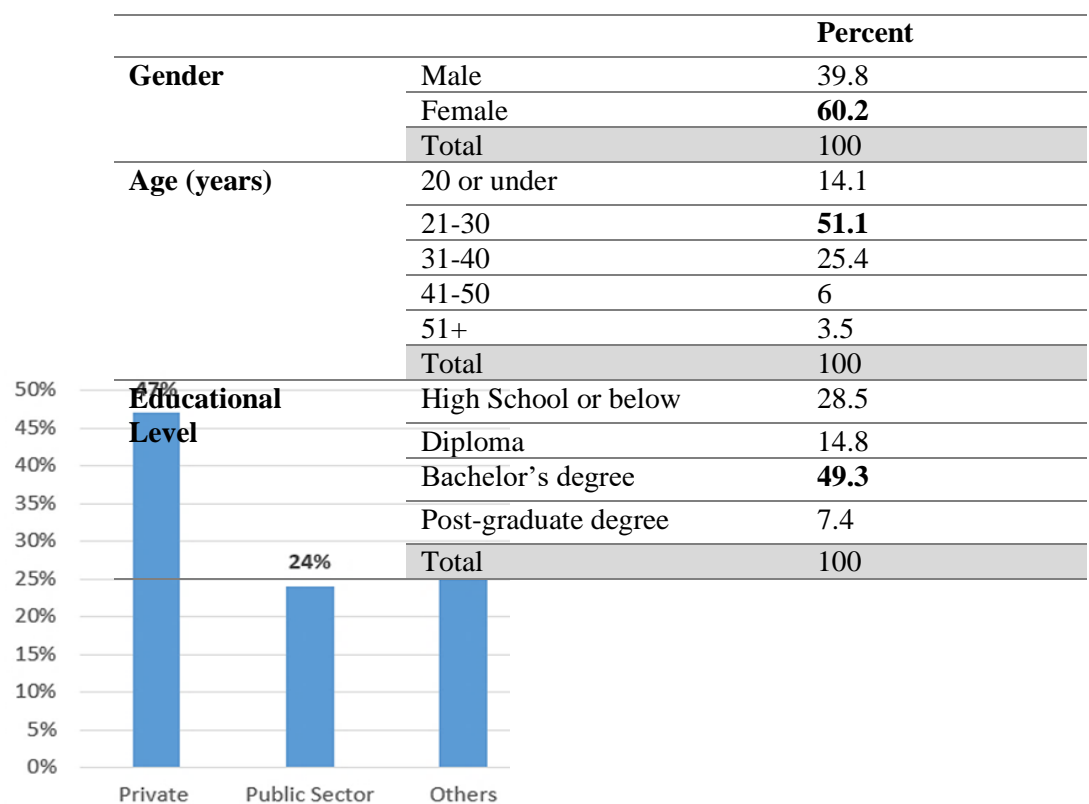

Figure 1 Respondents demographic data

The level of individual awareness and their perceptions with respect to OGD were indeed assessed in this study. Table 4 displays the percentage of population who heard about OGD. Interestingly, 45% of population knows about the term OGD. Besides being aware of OGD, the usage was undeniably considered as an important question that reflects the utilization of OGD by the public sector. Almost 70% of respondents did not use OGD at all. Only 30% of the respondents have attempted to use OGD in some way as shown in Table 4. Table 4 also shows the distribution of respondents who attempted to use OGD based on their working sector. Private and public sector workers occupied the largest proportion of OGD users in Bahrain as they accounted for almost two thirds of OGD users. Unfortunately, Entrepreneurs and other non-government, non-private organizations such as civil society and foundation charity exhibit an almost total lack of use of OGD in Kingdom of Bahrain.

Table 4 Awareness and usage of OGD

| Knowledge about OGD | |
|---|---|
| Yes | 45% |
| No | 55% |
| Using OGD | |
| Yes | 30.3% |
| No | 67.7% |
| **Sectors who use OGD** | |
| Foundation/Charity organization | 1.0% |
| Entrepreneurship | 1.0% |
| Media | 2.5% |
| Academia/Research | 2.0% |
| Student | 3.0% |
| Public Sector | 10.0% |
| Private Sector | 10.0% |
| | |

Additionally, the association between the population who heard about OGD and the population who used it was examined by utilizing a Chi-Square Test of Association. Table 5 confirms that there is a significant association between being aware of OGD and using OGD (X2(1) = 139.932, p < .001).

Table 5 Test of association between being knowledgeable about OGD and using OGD

| | | Have you ever used OGD | | |
|---|---|---|---|---|
| | | Yes | No | Total |
| Have you ever heard about "Open Government Data"? | Yes | 84 | 43 (22%) | 127 |
| | No | 2 | 155 (78%) | 157 |
| Total | | 86 | 198(100%) | 284 |
| **Chi-Square Test** | | | | |
| | Value | df | Asymptotic Significance (2-sided) | Exact Sig. (2-sided) |
| Pearson Chi-Square | 139.932[a] | 1 | 0.000 | |
| Continuity Correction[b] | 136.876 | 1 | 0.000 | |
| Likelihood Ratio | 164.306 | 1 | 0.000 | |

a: 0 cells (0.0%) have expected count less than 5. The minimum expected count is 38.46.

Table 6 Types of OGD required by residents

| Rank | OGD Type | Very Important OGD Type Percentage |
|------|----------|-------------------------------------|
| 1 | Education (e.g. school performance, educational resources, etc.) | 69% |
| 2 | Security and justice (e.g. crime data) | 65% |
| 3 | Health and welfare (e.g. public health inspections, hospitals performance, etc.) | 65% |
| 4 | Legal, Politics and Policies (e.g. laws, official proceedings, bulletins, election results, etc.) | 65% |
| 5 | Government Operations (e.g. detailed national budget, Spending, taxes distribution, contracts, elections etc.) | 64% |
| 6 | Housing & Real estate | 64% |
| 7 | Business (trade, import& export etc.) | 62% |
| 8 | Tourism | 62% |
| 9 | Transportation (e.g. information about roads and public transportation) | 61% |
| 10 | Natural resources and environment (e.g. Pollution, etc.) | 60% |
| 11 | International/Global Development | 60% |
| 12 | Economics, Finance, Insurance | 60% |
| 13 | Statistics (e.g. socioeconomic and demographic information) | 59% |
| 14 | Climate | 58% |
| 15 | Geographical/Geo-spatial/Mapping Data (e.g. maps, points of interest, etc.) | 58% |
| 16 | Demographics and Social (Statistics, marriage and divorce, Labor force, etc.) | 58% |
| 17 | Consumer | 57% |
| 18 | Science and Research | 55% |
| 19 | Sports | 54% |
| 20 | Manufacturing | 54% |
| 21 | Agriculture | 51% |

The determination of people's requirements regarding what Open Government Data they really need was also identified in the current study. Twenty-one different categories of OGD were given and people were asked to rank the importance level of each category according to their opinion. As shown in Table 6, all of the items received above 50% ranking as being considered very important to make them open and available to the public. However, the highest ranked categories include "Educational Data" 69 %, "Security and justice" 65%, "Health and welfare" 65%, "Legal" 65%, "Government Operations" 64%, and "Housing and real-estate" 64%. In contrast, "Agriculture" was ranked as the lowest category that responds are interested in 51%. Figure 2 on the other hand, displays the purpose of using Open Government data by the respondents in Kingdom of Bahrain. The main two purposes were "For daily operation in work" and "For Commercial business use" 20% and 17%, respectively. Other purposes include "To perform statistical analysis" 13%, "For data linking", 13% and "To write academic publications" 13%.

Moreover, an important element in the assessment of respondents' awareness and perceptions towards OGD is to identify whether there is a demand for certain OGD that is not yet available to the public. Unexpectedly, the majority of respondents (67%) do think that all of the required data is available at the Kingdom of Bahrain's Open Government Data portal as shown in Figure 3.

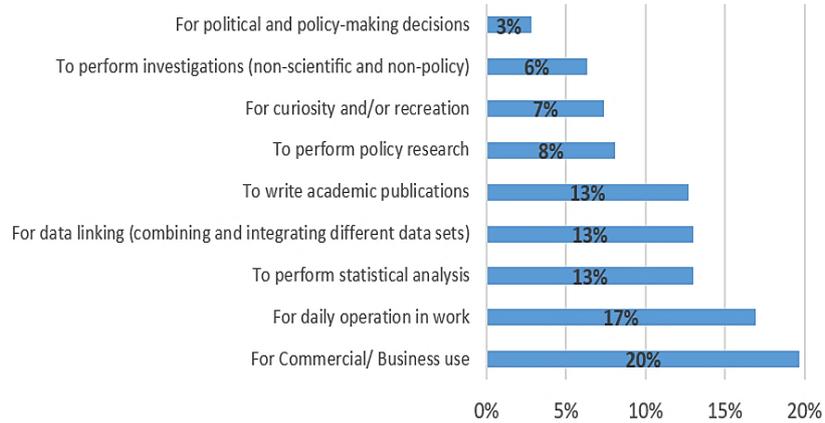

Figure 2 Purpose of using Open Government data by the respondents

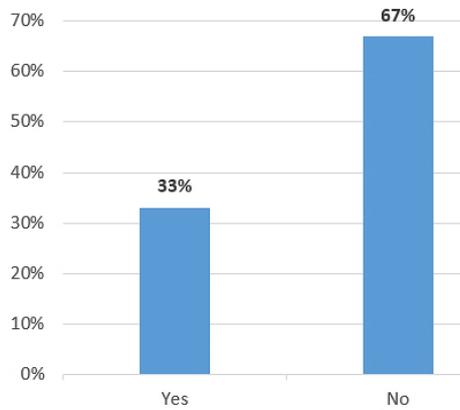

Figure 3 Respondents' opinion measuring shortage of OGD data provided by government

However, since this question has been answered by all of the population who either used or not used OGD, the results could be misleading. Table 7 shows the association between the use of OGD and the perception of respondents to weather there is a demand for certain Open Government data but unfortunately not yet available at OGD portal. It has been found that there is a significant association between the two variables ($X2(1) = 106.118$, $p < .001$). Thus, the people who used OGD do believe that there is a demand for OGD that has not yet been published by the government and vice versa. In statistical terms, 77% of respondents who used OGD admitted that OGD portal is not comprehensive and lacks the required data. In contrast, almost 86% of respondents who never used OGD believed that OGD portal is comprehensive and all of the required data could be found at the portal.

Table 7 Test of association between OGD users and lack of certain (in demand) datasets

| | | Do you think there is a demand for certain Open Government data but unfortunately it is not yet available at the national Open Government Data Portal? | | |
|---|---|---|---|---|
| | | Yes | No | Total |
| Have you ever used "Open Government Data"? | Yes | 66 | 20 | 86 |
| | No | 28 | 170 | 198 |
| Total | | 94 | 190 | 284 |
| **Chi-Square Test** | | | | |
| | Value | df | Asymptotic Significance (2-sided) | Exact Sig. (2-sided) |
| Pearson Chi-Square | 106.118[a] | 1 | 0.000 | |
| Continuity Correction[b] | 103.309 | 1 | 0.000 | |
| Likelihood Ratio | 105.946 | 1 | 0.000 | |

a. 0 cells (0.0%) have expected count less than 5. The minimum expected count is 28.46.

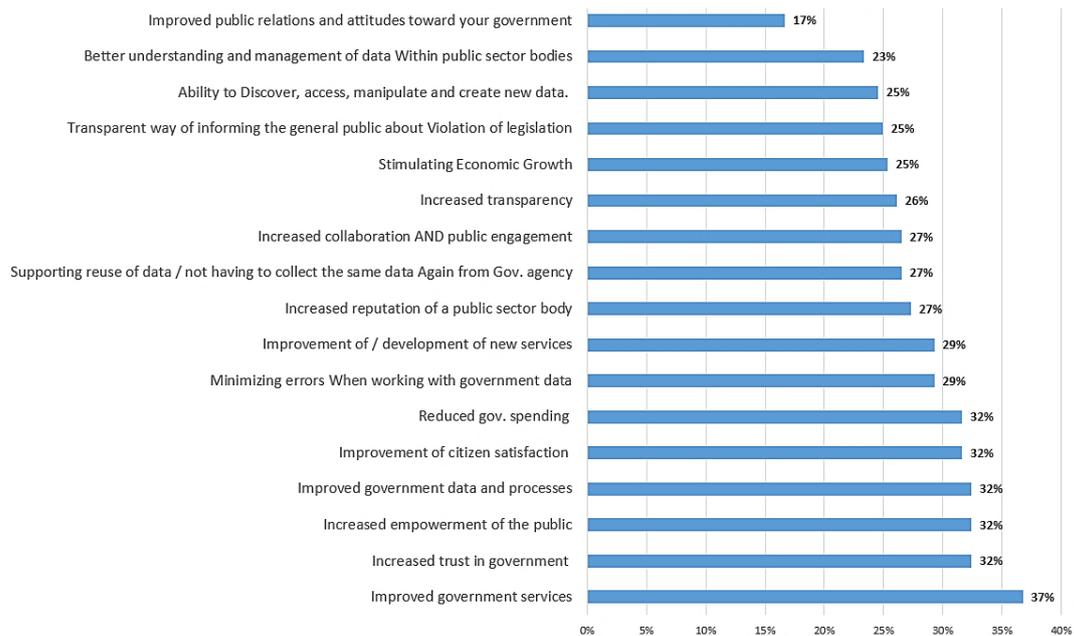

Figure 4 Respondents' perceived benefits of OGD

As with any new technology or concept, the use of OGD entails many benefits as well as barriers. The results pertaining to respondents' perceived benefits and barriers of using and utilizing open government data will be demonstrated next. With respect to the perceived benefits of OGD,

Figure 4 presents the rating of 17 mentioned benefits of OGD. The highest ranked benefits are "Improved government services", "increased trust in government", "increased empowerment to the public", "Improved Government data and processes" and "Improved citizen satisfaction" as they scored 37%,32%,32%,32%,32% respectively.

With respect to barriers, two categories of barriers were presented to respondents. The first category involved all barriers that could affect the use of Open Government Data by the people living in Kingdom of Bahrain. This kind of barriers is also known as use barriers or adoption barriers. Figure 5 shows the rating of barriers according to the opinions of respondents. The highest perceived barriers are: "Data is not up-to-date" (38%) and "Data is difficult to find" (38%). Other high rated barriers include: "Lack of support or helpdesk", "Lack of skills required to find and use the data", "Difficult to use OGD Portal" and "Data is not specific enough (data is aggregated) "as they scored 29%,26%,23%,23% respectively.

In the second category of barriers, respondents were asked to rate all of the barriers – based on their own opinion – that they believe could be currently preventing the government from opening more data to the public. This kind of barriers is known as Government OGD publication Barriers. Figure 6 shows that "Legal (e.g. privacy issues)" 72%, "Data availability (e.g. data is of bad quality or non-existent)" 71%, "Technological (e.g. lack of adequate infrastructure)" 70% are the greatest barriers that users believe in inhibiting the government from opening OGD to the public.

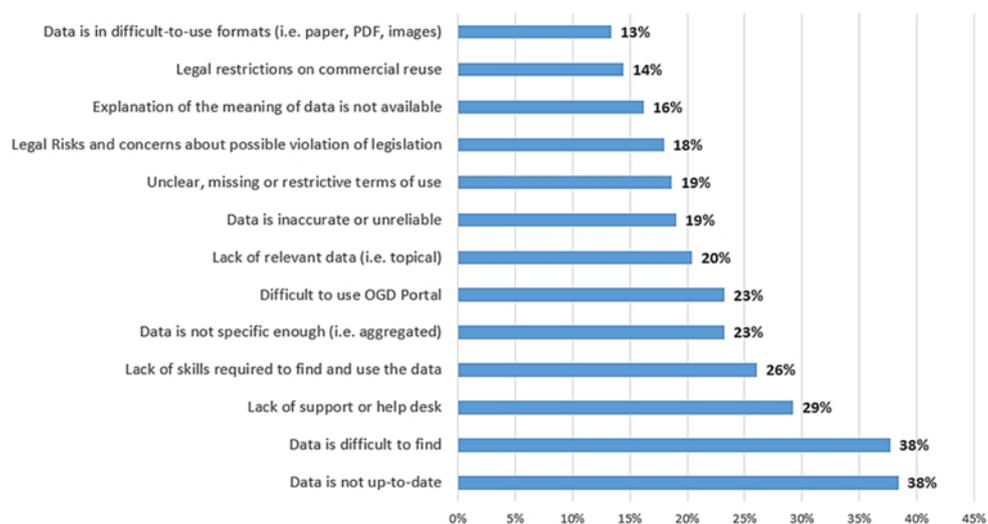

Figure 5 Respondents' perceived OGD use barriers

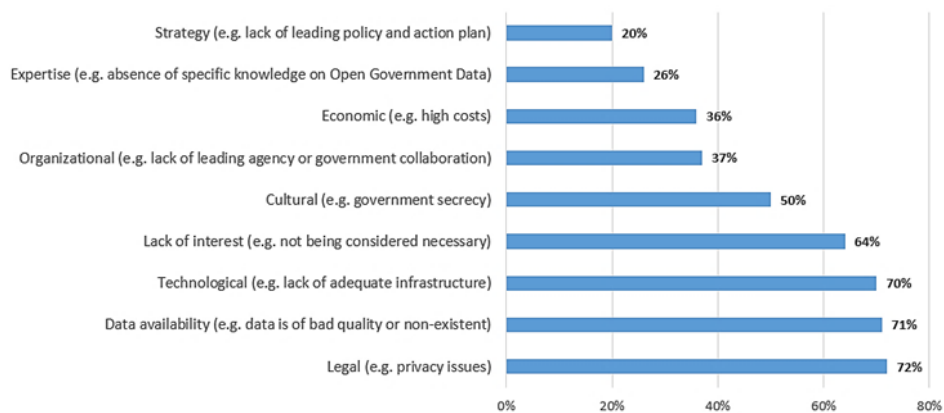

Figure 6 OGD Barriers to the government from people's perspective

## 5. RESEARCH DISCUSSION

The descriptive results can be used to establish a common picture regarding the awareness level of population towards OGD in Kingdom of Bahrain. Surprisingly, results confirm that there is a moderate level of awareness among the people living in Kingdom of Bahrain with respect to OGD. Almost half of population are -to some extent- aware of the term Open Government Data. Since OGD is relatively a new concept [47][48], being aware of it is an indication that the people in Kingdom of Bahrain are more technology-oriented community. This interpretation is also supported by a recent study done by Al-Ammary et al.[49] which confirms the existence of high level of e-government services awareness among the population living in Kingdom of Bahrain.

The source of user's awareness towards this concept could be either local efforts directed towards promoting OGD or some other ways such as international OGD promotion campaigns and news agencies. One main source of user's awareness in kingdom of Bahrain can be implicitly identified by looking at the results which indicates that the work environment necessitates for the workers to be knowledgeable about this term. Additionally, Alanazi and Chatfield [41] reported that internet penetration rate has a partial effect on open data practice. Thus, higher internet penetration in kingdom of Bahrain[41] can provide a partial reasoning as well to the moderate level of OGD awareness among the people living in Kingdom of Bahrain. It is important as well not to ignore that almost the other half of respondents did not hear at all about this concept. This indicates that there is a need for the government to raise OGD awareness to higher levels among the people in kingdom of Bahrain. In line with this, Saxena [44] explicitly stated that promoting OGD initiative among the people is essential for improving the implementation of OGD in GCC countries. Possible means for raising the awareness include sponsoring media and technical hackathons. Other means include utilizing social media channels as they are effective tools for increasing OGD awareness among the public[44], [50].

In contrast to the moderate level of OGD awareness among the population, the findings show that the actual use of OGD in Kingdom of Bahrain is Low, as almost 70% of the population did not use OGD at all. Several findings can be used to illustrate the reasoning of this low percentage. First, the results shown in Table 5 confirm the existence of strong association between being knowledgeable about OGD and using OGD. Thus, the people who did not hear about OGD did not use it and vice versa. In statistical terms, 78% of the people who did not use OGD are actually unaware of the meaning or existence of OGD. This indicates that in the context of Bahrain, awareness plays an important role in promoting the use of OGD. With this regard, Al-Ammary et al. [49] showed the existence of a strong relationship between the awareness of e-government services and the actual use of such services. Besides awareness, other factors also exist which can affect the remaining 22% of population - who are aware about OGD, but never attempted to use it. For example, the believe that OGD may not bring benefits to the public can hider and stop motivating the population from using OGD. Additionally, OGD use barriers can contribute to prevent the population who knows about OGD from using it. Such use barriers include difficulty in finding the data, lack of OGD support or helpdesk from the government side and lack of skills required to find Open Government Data. Thus, low level of OGD use in kingdom of Bahrain can be reasoned firstly to OGD awareness and secondly to the existence of many OGD use barriers and lack of full knowledge about the benefits of OGD which prevent the population from using the data.

With respect to those who used OGD which are 30% of population, it is important to clarify that this questionnaire did not ask for the frequency of usage, hence a user could be using OGD for only one time or for many times. Therefore, this percentage should be viewed carefully as it does not mean that users are successfully used and benefited from OGD. Instead, it refers to the attempt of using OGD regardless of the frequency of usage. Additionally, this percentage could also cover the use of some types of OGD data that are indirectly owned by the government and provided by subsidiaries, agencies or private organizations as illustrated by Attard et al. [5] . Such OGD include data about transportation, climate and traffic congestion. For example, Maps applications provided by non-government organizations such as google maps utilize some of OGD data that is owned/provided by the Government. Hence, a user of such applications would consider him/herself as a user of OGD.

Low level of OGD use means that OGD initiative in kingdom of Bahrain still lacks an important and essential element to make it successful, which is the actual use of OGD. As mentioned in the previous section, unless OGD is used, there is no benefit from the data itself [5]. The results of Open Data Barometer [15], [51] clearly show that there is an absence of any obvious political, social or economic impacts from OGD in Kingdom of Bahrain. Since the impact of OGD is tied directly to the actual utilization of OGD by the users, results of Open Data Barometer confirms and comes in line with the results of this study , ie: OGD use in Bahrain is low.

Low level of OGD use is also affecting the perception of people living in Kingdom of Bahrain with respect to whether there is a shortage of OGD data provided by the government to the public. The results provided in Figure 3 and Table 7 clearly shows that there is a clear shortage of OGD Data required by respondents. However, only the users who tried OGD are able to state with confidence that there is a shortage of OGD data. Because other respondents did not use OGD, they mistakenly believed that all of OGD data is available and provided by the Government. This result is a clear indication ( as mentioned above) of the need to raise the level of awareness and improve OGD use among the population in Kingdom of Bahrain.

The identification of People's OGD Requirements is necessary to help the government agencies in determining what to publish and when. With this regard, the ranking of 19 different categories of OGD by respondents can be used to draw several reflections related to the actual demand of OGD in Bahrain. First, the results of ranking show clearly that respondents are interested in all types of OGD. This is concluded from the fact that all the presented categories have scored above 50% as being very important to the people in Kingdom of Bahrain. This means that the government can publish any type of OGD within the identified categories and be confident that there will be a demand for it by the people living in kingdom of Bahrain.

Second, besides the fact that all of OGD is important to respondents, results prove that some data got more preference than the others. Such as education, security, justice and health. This prioritization of OGD types could be reasoned to either that such data is really considered of high importance to the public at all times, or the prioritization is affected by the existing circumstances or events that are occurring in the current time. The latter case is evident when comparing the highly rated OGD datasets with the current circumstances in Bahrain. For example, the educational system in Bahrain is currently subjecting to many improvement projects and laws [52] . Such improvements include the introduction of new college specialties and universities, the movement towards smart schools and the obvious change in subject teaching and information delivery to the students. Another example is the Health data. New healthcare reforms are being introduced in the current time in Kingdom of Bahrain [53] and [54]. One of the reforms is the national health information system (I-Seha) which is already being deployed in many health institutions and aims to provide highly efficient health services to the public. Another recent reform is the Bahrain's national health insurance project that are widely advertised and expected to be rolled out for implementation soon. All the above-mentioned projects are recent and touches the daily lives of people in Kingdom of Bahrain. Thus, such examples illustrate some unusual reformation / improvement activities by the government which certainly affects the people and thereby might affect their preferences towards what data they really need to know about. In this situation, the government must be prepared in advance to meet the data requirements of people whenever there is a major initiative/ project being undertaken. By doing so, the success of new reforms or projects will be certainly fostered in the sense that the questions and doubts of people will find useful answers from the published Open Government Data.

Benefits and Barriers of OGD are also important elements that must be considered when implementing the OGD initiative. This section will discuss and interpret the results of the perceived benefits and barriers presented earlier. With respect to the perceived benefits, it is noticeable that the rating of benefits by the people follows – to some extend – a logical sequence in which respondents believed more in the benefits that can be categorized as quick or noticeable gains compared to other benefits that require wide OGD adoption and relatively longer implementation and use time. For example, respondents perceived highly the benefit that OGD will improve government services, which as a result could contribute to the development of new services – a less ranked benefit. Another example is the following highly rated benefits (increased trust in government, increased empowerment of public, improved data and services, improvement of citizen satisfaction and reduced Government spending) can be considered essential and necessary to attain the less ranked benefits (increased transparency and stimulating economic growth) which are usually noticeable in the long term. Moreover, this raking of OGD benefits can provide additional useful insights as well. The government can utilize upon these results to promote OGD by advertising the benefits that the majority of population believe in. For example, when launching new datasets or OGD mobile application, the government can promote its activities by highlighting that OGD will empower the public and improve their satisfaction. This promotion of the benefits will motivate the population to attempt using OGD.

Additionally, the application of perceived benefits can be applied in a two-way direction. Specifically speaking, (I) Government provides specific OGD and communicates the benefits to the public, (ii) the public uses the provided valued OGD and realizes the promised benefits and as a result of this, the government will gain some additional benefits as well. For example, people in Kingdom of Bahrain believed that OGD could increase trust in Government. Hence, the government can utilize this point to provide certain OGD data such as the exact distribution of collected taxes (where taxed are spent) or data that shows the effect of traffic fines increment on the total number of accidents occurring in kingdom of Bahrain. By doing so, people will feel empowered and will trust more the actions of the government. On the other way, once the level of Government trust by citizen is improved, the government will gain a bonus benefit in the sense that it will become easy for it to issue, justify new regulations, and get the support from the public. These two-way processes can ultimately contribute in reaching the other benefits.

OGD barriers from people's perspective are also identified in this study from two different viewpoints: Citizens' perceived OGD use barriers and citizens' perceived OGD implementation barriers. With respect to OGD use barriers, it is assured that there is an issue with OGD data discovery process in kingdom of Bahrain. Specifically, there are issues with the OGD Portal as the highly considered barriers were "data is difficult to find" and "data is not up-to- data". Such issues can be related to functionality of the portal and the availability of the required data. This interpretation comes in-line with several studies and evaluations that indicate the poor quality of OGD portal. For example, results of Martín et al. [13] study showed that the quality of OGD portal in kingdom of Bahrain is low. Specifically, timeliness of data was an obvious issue (data is not up to date barrier). However, their results showed that OGD portal is equipped with many functional features that could help in finding the data. Such results contradict what people believe (Data is difficult to find). The explanation of this can be recognized easily once a person tries to use OGD portal; Useful functionalities exist in Bahrain's OGD portal but they do not work properly as noted by me and other scholars as well. Many studies confirm the low quality of OGD portal such as Saxena [22] who mentioned this issue when she tried to utilize the search functionality but the results were far away from what was required. Thus, OGD portal is suffering from many issues including functional issues that led the respondents to highly rate "Data is difficult to find barrier".

Moreover, the scoring of OGD use barriers implicitly reflect the fact of low OGD use in kingdom of Bahrain. Since many scholars have already identified the existence of data and metadata quality issues in kingdom of Bahrain OGD portal [21], perceiving barriers such as "explanation of the meaning of the data is not available - Metadata quality issue", "Data is in difficult to use formats" by only small percentage of respondents indicate that the others have never tried to use OGD. This is because such less scored barriers cannot be fully comprehended unless the user finds the required data and tries to use it. At that time, the user will possibly realize the existence of such barriers that scholars confirm its existence. Thus, this study confirms the existence of many barriers that are already identified in previous studies and evaluations. With regard to the perceived OGD initiative implementation barriers from people's perspectives, the findings demonstrate implicitly that respondents have trust and confidence in the leadership of their government. This is illustrated by the fact that the leadership-related aspects such as strategy, expertise and organizational barriers were not believed by respondents to be real implementation barriers in the context of Bahrain. Rather, the highly rated barriers were related to legal, data availability or technological aspects.

## 6. CONCLUSION

Based on the results of this study, government organizations are recommended to employ an effective awareness campaigns that target all expected OGD users: general citizens/residents, academia, entrepreneurs and IT developers. Such awareness campaigns could be established through the utilization of social media channels and organizing hackathons. It is important to note that before promoting OGD to any segment especially the public, certain quality datasets and evident use cases should be presented to the public as a way of validating the arguments of the government and increasing the motivation of people towards using and benefiting from OGD.

An action plan to improve the OGD portal functions, OGD dataset and metadata quality aspects should be developed and applied. In fact, a complete reform for the OGD portal has to be done to incorporate useful, general and OGD specific functionalities. All OGD Datasets should exhibit high quality, not varied quality, levels. In addition, high quality metadata information should also be provided with all OGD datasets.

Finally, specific nationwide legislation regarding OGD publication and use must be identified and enforced. Such legislation should include laws that state the right of citizens to know and use OGD. It should also address the security and privacy issues related to publishing and using OGD as well as the authorities and responsibilities of public bodies to make and enforce specific decisions regarding publishing OGD.

## ACKNOWLEDGEMENTS

The authors would like to thank Bahrain Information And e-Government Authority for their great and usual support especially in data collection from their various governmental partners.

## REFERENCES

[1]   W. Carrara, M. Nieuwenhuis, and H. Vollers, *Open Data Maturity in Europe 2016: Insights into the European State of Play*. Capgemini Consulting, 2016.

[2]   A. T. Chatfield and C. G. Reddick, "A longitudinal cross-sector analysis of open data portal service capability: The case of Australian local governments," *Gov. Inf. Q.*, vol. 34, pp. 231–243, 2017.

[3]   K. Jung and H. W. Park, "A semantic (TRIZ) network analysis of South Korea's _Open Public Data_ policy," *Gov. Inf. Q.*, vol. 32, pp. 353–358, 2015.

[4]   T. van den Broek, A. F. van Veenstra, and E. Folmer, "Walking the extra byte : A lifecycle model for linked open data," in *Pilot linked open data Nederland. Deel 2 De verdieping,* Bizzprint, 2013, pp. 94–110.

[5]   J. Attard, F. Orlandi, S. Scerri, and S. Auer, "A systematic review of open government data initiatives," *Gov. Inf. Q.*, vol. 32, pp. 399–418, 2015.

[6]   B. Ubaldi, *Open Government Data Towards Empirical Analysis of Open Government Data Initiatives*. 2013.

[7]   F. W. Donker and B. van Loenen, "How to assess the success of the open data ecosystem?," *Int. J. Digit. Earth*, vol. 10, pp. 284–306, 2016.

[8]   A. Zuiderwijk, M. Janssen, and C. Davis, "Innovation with open data: Essential elements of open data ecosystems," *Inf. Polity*, vol. 19, pp. 17–33, 2014.

[9]   S. S. Dawes, L. Vidiasova, and O. Parkhimovich, "Planning and designing open government data programs: An ecosystem approach," *Gov. Inf. Q.*, vol. 33, pp. 15–27, 2016.

[10]  S. Shekhar and V. Padmanabhan, *How to support the capacity of open data initiatives with assessment tools*. 2016.

[11]  H.-C. Yang, C. S. Lin, and P.-H. Yu, "Toward Automatic Assessment of the Categorization Structure of Open Data Portals," in *Multidisciplinary Social Networks Research: Second International Conference, MISNC 2015, Matsuyama, Japan, September 1-3, 2015. Proceedings*, L. Wang, S. Uesugi, I. H. Ting, K. Okuhara, and K. Wang, Eds. Springer Berlin Heidelberg, 2015, pp. 372–380.

[12]  J. Kučera, D. Chlapek, and M. Nečaský, "Open Government Data Catalogs: Current Approaches and Quality Perspective," Springer, Berlin, Heidelberg, 2013, pp. 152–166.

[13]  A. Martín, A. De Rosario, and M. Pérez, "An International Analysis of the Quality of Open Government Data Portals," *Soc. Sci. Comput. Rev.*, vol. 34, no. 3, pp. 298–311, 2015.

[14]  K. Reiche and E. Hofig, "Implementation of Metadata Quality Metrics and Application on Public Government Data," in *2013 IEEE 37th Annual Computer Software and Applications Conference Workshops (COMPSACW)*, IEEE, 2013, pp. 236–241.

[15]  O. D. Barometor, "Country Detail | Open Data Barometor." World Wide Web Foundation, 2015.

[16]  O. D. Barometer, "Open Data Barometer 3rd Edition - Global Report." World Wide Web Foundation, 2017.

[17]  O. K. Foundation, "Global Open Data index." Open Knowledge Foundation, 2015.

[18]  A. K. Katbi and J. Al-Ammary, "Open Government Data in Kingdom of Bahrain: Towards an Effective Implementation Framework," in *Advances in Intelligent Systems and Computing*, 2019, vol. 930, pp. 699–715, doi: 10.1007/978-3-030-16181-1_66.

[19]  I. Elbadawi, "The state of open government data in GCC countries," in *Proceedings of the 12th European Conference on eGovernment*, Institute of Public Governance and Management, 2012, pp. 193–200.

[20]  S. Saxena, "Open public data (OPD) and the Gulf Cooperation Council (GCC): challenges and prospects," *Contemp. Arab Aff.*, pp. 1–13, 2017.

[21]  "Bahrain Economic Vision 2030." [Online]. Available: https://bahrainedb.com/

[22]  P. Srinivasan, "The GovLab Index: Open Data - 2016 Edition - The Governance Lab @ NYU." 2017.

[23]  A. Al-Khouri, "Open Data: A Paradigm Shift in the Heart of Government," *J. Public Adm. Gov.*, vol. 4, p. 217, 2014.

[24]  A. Zuiderwijk, "Open data infrastructures: The design of an infrastructure to enhance the coordination of open data use," 2015.

[25]  J. Kucera, "Open Government Data," 2014.

[26]  T. Jetzek, M. Avital, and N. Bjørn-Andersen, "Generating sustainable value from open data in a sharing society," in *IFIP Advances in Information and Communication Technology*, 2014, vol. 429, pp. 62–82, doi: 10.1007/978-3-662-43459-8_5.

[27]  E. G. Martin and G. M. Begany, "Opening government health data to the public: benefits, challenges, and lessons learned from early innovators," *J. Am. Med. Informatics Assoc.*, pp. 345–351, 2016.

[28]  A. Zuiderwijk, R. Shinde, and M. Janssen, "Investigating the attainment of open government data objectives: Is there a mismatch between objectives and results?," *Int. Rev. Adm. Sci.*, p. 002085231773911, 2018.

[29]  U. Nations, *Open Government Data Assessment Methodology: Guide on OGD Action Planning for Sustainable Development*. 2015.

[30]  M. Janssen, Y. Charalabidis, and A. Zuiderwijk, "Benefits, Adoption Barriers and Myths of Open Data and Open Government," *Inf. Syst. Manag.*, vol. 29, pp. 258–268, 2012.

[31]  A. Zuiderwijk, M. Janssen, S. Choenni, R. Meijer, and R. S. Alibaks, "Socio-technical Impediments of Open Data," *Electron. J. e-Government*, vol. 10, pp. 156–172, 2012.

[32]  A. Zuiderwijk and M. Janssen, "Barriers and Development Directions for the Publication and Usage of Open Data: A Socio-Technical View," in *Open Government Opportunities and Challenges for Public Governance*, M. Gasc_-Hern_ndez, Ed. Springer, 2014, pp. 115–135.

[33]  J. Brugger, M. Fraefel, R. Riedl, H. Fehr, D. Schöeneck, and C. S. Weissbrod, "Current Barriers to Open Government Data Use and Visualization by Political Intermediaries," in *2016 Conference for E-Democracy and Open Government (CeDEM)*, 2016, pp. 219–229.

[34]    K. M. Chorley, "The challenges presented to records management by open government data in the public sector in England," *Rec. Manag. J.*, vol. 27, pp. 149–158, 2017.

[35]    G. Magalhães and C. Roseira, "Exploring the Barriers in the Commercial Use of Open Government Data," in *Proceedings of the 9th International Conference on Theory and Practice of Electronic Governance*, 2016, pp. 211–214.

[36]    P. Conradie and S. Choenni, "On the barriers for local government releasing open data," *Gov. Inf. Q.*, vol. 31, pp. S10–S17, 2014.

[37]    M. Najafabadi and L. F. Luna-Reyes, "Open Government Data Ecosystems: A Closed-Loop Perspective," in *50th Hawaii International Conference on System Sciences*, Hawaii, 2017, pp. 2711–2720.

[38]    L. Ding *et al.*, "TWC LOGD: A portal for linked open government data ecosystems," *Web Semant. Sci. Serv. Agents World Wide Web*, vol. 9, pp. 325–333, 2011.

[39]    L. Reggi and S. Dawes, "Open Government Data Ecosystems: Linking Transparency for Innovation with Transparency for Participation and Accountability," in *International Conference on Electronic Government and the Information Systems Perspective EGOVIS*, Springer International Publishing, 2016, pp. 74–86.

[40]    S. Martin, S. Turki, and S. Renault, "Open Data Ecosystems Introducing the Stimulator Function," in *Electronic Government and the Information Systems Perspective*, Springer International Publishing, 2017, pp. 49–63.

[41]    J. Alanazi and A. Chatfield, "Sharing Government-Owned Data with the Public: A Cross-Country Analysis of Open Data Practice in the Middle East," in *In Proceedings of Americas Conference on Information Systems*, 2012.

[42]    O. G. W. Group, "The 8 Principles of Open Government Data ." 2007.

[43]    N. Alromaih, H. Albassam, and H. Al-Khalifa, "A proposed checklist for the technical maturity of open government data: an application on GCC countries," in *Proceedings of the 18th International Conference on Information Integration and Web-based Applications and Services*, ACM, 2016, pp. 494–499.

[44]    S. Saxena, "Prospects of open government data (OGD) in facilitating the economic diversification of GCC region," *Inf. Learn. Sci.*, vol. 118, pp. 214–234, 2017.

[45]    M. Denscombe, *The good research guide*. McGraw-Hill/Open University Press, 2011.

[46]    M. Saunders, P. Lewis, and A. Thornhill, *Research methods for business students*. Pearson Education Limited, 2009.

[47]    J. Attard, F. Orlandi, and S. Auer, "Value Creation on Open Government Data," in *49th Hawaii International Conference on System Sciences (HICSS), 2016*, IEEE, 2016, pp. 2605–2614.

[48]    T.-M. Yang and Y.-J. Wu, "Examining the socio-technical determinants influencing government agencies' open data publication: A study in Taiwan," *Gov. Inf. Q.*, vol. 33, pp. 378–392, 2016.

[49]    J. Al-Ammary, R. AL-Kaabi, A. Al-Soufi, H. Ali, A. AlRayes, and M. Aljawder, "Assessment of e-Government Services from the Supply Side to the Demand Side: Towards Better e-Services in the Kingdom of Bahrain," *Int. J. Econ. Res.*, vol. 14, 2017.

[50]    G. Lee and Y. H. Kwak, "An Open Government Maturity Model for social media-based public engagement," *Gov. Inf. Q.*, vol. 29, pp. 492–503, 2012.

[51]    O. D. Barometer, "Country Detail | Open Data Barometer." World Wide Web Foundation, 2016.

[52]    M. Education, *National Higher Education Strategy: putting higher education at the heart of the nation*. 2014.

[53]    M. Health, *Health Improvement Strategy 2015-2018*. 2015.

[54]    O. B. Group, "Changes to Bahrain's health care sector as it adapts to new conditions." Oxford Business Group, 2016.

## Authors

**Abdulkarim Katbi** is a Ph.D. student at University of Bahrain. He received his BSc. in Information Systems from University of Bahrain in 2011 and MSc. in Information Technology from University of Bahrain in 2019. He is currently working as a computer systems and network administrator at King Hamad University Hospital, kingdom of Bahrain. His areas of interest are in Open Data, Strategic Information Systems, and Software evolution and visualization.

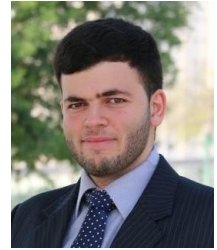

**Jaflah Hassan Al-Ammary** is assistant professor in Department of Information Systems, College of IT, University of Bahrain. She holds PhD from the University of Murdoch (Australia). Al-Ammary's research interest focuses on IS strategies and Applications focus more on Strategic alignment, Knowledge Management, and educational technology and E-learning. Al-Ammary has many publications on KM-strategic alignment and the strategic role of the Information systems and SISP at the organization of Bahrain. Al-Ammary is currently part of a national project for Information and e-Government Authority, Bahrain; to assess the customer satisfaction index (BHSCI). She is also collaborating with Ministry of Education Regional Center for ICT to enhance the research in Educational Technology and innovation.

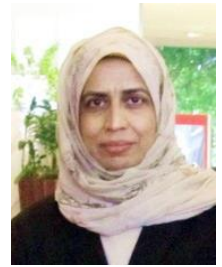

**Ali li AlSoufi** is the Dean of ICT at British University of Bahrain, an x-Associate professor of Information Systems at University of Bahrain. He has earned his PhD in computer science in 1994 from Nottingham University, UK. Worked for Bahrain Telecom Co for 8 years as a Senior Manager Application Programme, where he overlooked number of mega IS Application projects. Worked at Arab Open University as Director of IT program and Assistant Director for Business Development during 2007-2010. He is a PT consultant in Bahrain Information and e-Government Authority (iGA) in the area of Enterprise Architecture and Strategic Planning. He is an active member of the Bahrain National ICT Governance Committee. His specializations is Strategic IT Planning and Governance, IT project management, Enterprise Architecture and Information Systems in Organization.

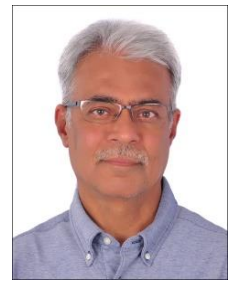